\documentstyle[ltwol,epsf,rotate]{article}

\def\be{\begin{equation}}
\def\ee{\end{equation}}
\def\bea{\begin{eqnarray}}
\def\eea{\end{eqnarray}}
 
\begin{document}


\begin{titlepage}
 
\begin{flushright}
CERN-TH/98-296\\
hep-ph/9809302
\end{flushright}
 
\vspace{2.5cm}
 
\begin{center}
\Large\bf Towards the Extraction of the CKM Angle \boldmath$\gamma$\unboldmath
\end{center}

\vspace{1.8cm}
 
\begin{center}
Robert Fleischer\\
{\sl Theory Division, CERN, CH-1211 Geneva 23, Switzerland}
\end{center}
 
\vspace{2.3cm}
 
\begin{center}
{\bf Abstract}\\[0.3cm]
\parbox{11cm}{
The determination of the angle $\gamma$ of the unitarity triangle of the 
CKM matrix is regarded as a challenge for future $B$-physics experiments.
In this context, the decays $B^\pm\to \pi^\pm K$ and $B_d\to\pi^\mp K^\pm$,
which were observed by the CLEO collaboration last year, received a lot
of interest in the literature. After a general parametrization of their 
decay amplitudes, strategies to constrain and determine the CKM angle 
$\gamma$ with the help of the corresponding observables are reviewed. 
The theoretical accuracy of these methods is limited by certain 
rescattering and electroweak penguin effects. It is emphasized that the 
rescattering processes can be included in the bounds on $\gamma$ by using 
additional experimental information on $B^\pm\to K^\pm K$ decays, and steps 
towards the control of electroweak penguins are pointed out. Moreover,
strategies to probe the CKM angle $\gamma$ with the help of
$B_s\to K\overline{K}$ decays are briefly discussed.}
\end{center}
 
\vspace{2.2cm}
 
\begin{center}
{\sl Invited talk given at the\\
XXIX International Conference on High Energy Physics -- ICHEP '98,\\
Vancouver, B.C., Canada, 23--29 July 1998\\ 
To appear in the Proceedings}
\end{center}
 
\vspace{1.5cm}
 
\vfil
\noindent
CERN-TH/98-296\\
September 1998
 
\end{titlepage}
 
\thispagestyle{empty}
\vbox{}
\newpage
 
\setcounter{page}{0}
 

\title{TOWARDS THE EXTRACTION OF THE CKM ANGLE \boldmath$\gamma$\unboldmath}

\author{ROBERT FLEISCHER}
 
\address{Theory Division, CERN, CH-1211 Geneva 23, Switzerland\\
E-mail: Robert.Fleischer@cern.ch}


\twocolumn[\maketitle\abstracts{ 
The determination of the angle $\gamma$ of the unitarity triangle of the 
CKM matrix is regarded as a challenge for future $B$-physics experiments.
In this context, the decays $B^\pm\to \pi^\pm K$ and $B_d\to\pi^\mp K^\pm$,
which were observed by the CLEO collaboration last year, received a lot
of interest in the literature. After a general parametrization of their 
decay amplitudes, strategies to constrain and determine the CKM angle 
$\gamma$ with the help of the corresponding observables are reviewed. 
The theoretical accuracy of these methods is limited by certain 
rescattering and electroweak penguin effects. It is emphasized that the 
rescattering processes can be included in the bounds on $\gamma$ by using 
additional experimental information on $B^\pm\to K^\pm K$ decays, and steps 
towards the control of electroweak penguins are pointed out. Moreover,
strategies to probe the CKM angle $\gamma$ with the help of
$B_s\to K\overline{K}$ decays are briefly discussed.}]

\section{Introduction}\label{intro}
Among the central targets of the $B$-factories, which will start operating in
the near future, is the direct measurement of the three angles $\alpha$, 
$\beta$ and $\gamma$ of the usual, non-squashed, unitarity triangle of the 
Cabibbo--Kobayashi--Maskawa matrix (CKM matrix). From an experimental point 
of view, the determination of the angle $\gamma$ is \mbox{particularly}  
challenging, although there are several strategies on the market, 
allowing -- at least in principle -- a theoretically clean extraction of 
$\gamma$ (for a review, see for instance Ref.~1). 

In order to obtain direct information on this angle in an experimentally 
feasible way, the decays $B^+\to\pi^+ K^0$, $B^0_d\to\pi^-K^+$ and their 
charge conjugates appear very 
promising.\,\cite{PAPIII}$^{\mbox{-}}$\cite{wuegai}\, Last year, 
the CLEO collaboration reported the observation of several exclusive 
$B$-meson decays into two light pseudoscalar mesons, including also these 
modes.\,\cite{cleo}\, So far, only results for the combined branching ratios
\begin{eqnarray}
\lefteqn{\mbox{BR}(B^\pm\to\pi^\pm K)\equiv}\nonumber\\
&&\frac{1}{2}\left[\mbox{BR}(B^+\to\pi^+K^0)+
\mbox{BR}(B^-\to\pi^-\overline{K^0})\right]\\
\lefteqn{\mbox{BR}(B_d\to\pi^\mp K^\pm)\equiv}\nonumber\\
&&\frac{1}{2}\left[\mbox{BR}(B^0_d\to\pi^-K^+)+
\mbox{BR}(\overline{B^0_d}\to\pi^+K^-)\right]
\end{eqnarray}
have been published, with values at the $10^{-5}$ level and large experimental 
uncertainties. 

A particularly interesting situation arises if the ratio
\begin{equation}\label{Def-R}
R\equiv\frac{\mbox{BR}(B_d\to\pi^\mp K^\pm)}{\mbox{BR}(B^\pm\to\pi^\pm K)}
\end{equation}
is found to be smaller than 1. In this case, the following allowed range
for $\gamma$ is implied:\,\cite{fm2}
\begin{equation}\label{gamma-bound1}
0^\circ\leq\gamma\leq\gamma_0\quad\lor\quad180^\circ-\gamma_0\leq\gamma
\leq180^\circ,
\end{equation}
where $\gamma_0$ is given by
\begin{equation}\label{gam0}
\gamma_0=\arccos(\sqrt{1-R})\,.
\end{equation} 
Unfortunately, the present data do not yet provide a definite answer to the 
question of whether $R<1$. The results reported by the CLEO collaboration 
last year give $R=0.65\pm0.40$,\,\cite{cleo} whereas an updated analysis,
which was presented at this conference, yields 
$R=1.0\pm0.4$.\,\cite{newCLEO}\, Since (\ref{gamma-bound1}) is 
complementary to the presently allowed range of 
$41^\circ\mathrel{\hbox{\rlap{\hbox{\lower4pt\hbox{$\sim$}}}\hbox{$<$}}}\gamma
\mathrel{\hbox{\rlap{\hbox{\lower4pt\hbox{$\sim$}}}\hbox{$<$}}}
134^\circ$ arising from the usual fits of the unitarity 
triangle,\,\cite{burasHF97} this bound would be of particular phenomenological
interest (for a detailed study, see Ref.~9). It relies on the 
following three assumptions:
\begin{itemize}
\item[i)] $SU(2)$ isospin symmetry can be used to derive relations between 
the $B^+\to\pi^+ K^0$ and $B^0_d\to\pi^-K^+$ QCD penguin amplitudes.
\item[ii)] There is no non-trivial CP-violating weak phase present in the
$B^+\to\pi^+ K^0$ decay amplitude.
\item[iii)] Electroweak (EW) penguins play a negligible role in the decays
$B^+\to\pi^+ K^0$ and $B^0_d\to\pi^-K^+$.
\end{itemize}
Whereas (i) is on solid theoretical ground, provided the ``tree'' 
and ``penguin'' amplitudes of the $B\to\pi K$ decays are defined 
properly,\,\cite{bfm} (ii) may be affected by rescattering 
processes of the kind $B^+\to\{\pi^0K^+\}\to
\pi^+K^0$.\,\cite{FSI}$^{\mbox{-}}$\cite{atso}\,
As for (iii), EW penguins may also play a more important role than is 
indicated by simple model calculations.\,\cite{groro,neubert}\, Consequently, 
in the presence of large rescattering and EW penguin effects, strategies 
more sophisticated\,\cite{defan,rf-FSI} than the ``na\"\i ve'' bounds 
sketched above are needed to probe the CKM angle $\gamma$ with $B\to\pi K$ 
decays. Before turning to these methods, let us first have a look at 
the corresponding decay amplitudes.

\boldmath
\section{The General Description of $B^\pm\to\pi^\pm K$ and
$B_d\to\pi^\mp K^\pm$ within the Standard Model}
\unboldmath
Within the framework of the Standard Model, the most important contributions 
to the decays $B^+\to\pi^+K^0$ and $B_d^0\to\pi^-K^+$ arise from QCD penguin 
topologies. The $B\to\pi K$ decay amplitudes can be expressed as follows:
\begin{eqnarray}
\lefteqn{A(B^+\to\pi^+K^0)=\lambda^{(s)}_u(P_u+P_{\rm ew}^u+
{\cal A})}\nonumber\\
&&\qquad+\lambda^{(s)}_c(P_c+P_{\rm ew}^c)+\lambda^{(s)}_t(P_t+
P_{\rm ew}^t)\label{ampl1}\\
\lefteqn{A(B^0_d\to\pi^-K^+)=-\left[\lambda^{(s)}_u(\tilde P_u+\tilde 
P_{\rm ew}^u+\tilde{\cal T}\,)\right.}\nonumber\\
&&\qquad\left.+\lambda^{(s)}_c(\tilde P_c+\tilde P_{\rm ew}^c)+
\lambda^{(s)}_t(\tilde P_t+\tilde P_{\rm ew}^t)\right],\label{ampl2}
\end{eqnarray}
where $P_q$, $\tilde P_q$ and $P_{\rm ew}^q$, $\tilde P_{\rm ew}^q$ denote 
contributions from QCD and electroweak penguin topologies with internal $q$ 
quarks $(q\in\{u,c,t\})$, respectively, ${\cal A}$ is related to annihilation 
topologies, $\tilde{\cal T}$ is due to colour-allowed $\bar b\to\bar uu\bar s$ 
tree-diagram-like topologies, and $\lambda^{(s)}_q\equiv V_{qs}V_{qb}^\ast$
are the usual CKM factors. Because of the tiny ratio
$|\lambda^{(s)}_u/\lambda^{(s)}_t|\approx0.02$, the QCD penguins play the 
dominant role in Eqs.\ (\ref{ampl1}) and (\ref{ampl2}), despite their
loop suppression. 

Making use of the unitarity of the CKM matrix and applying the Wolfenstein 
parametrization\,\cite{wolf} yields
\begin{equation}\label{Bpampl}
A(B^+\to\pi^+K^0)=-\left(1-\frac{\lambda^2}{2}\right)\lambda^2A\left[
1+\rho\,e^{i\theta}e^{i\gamma}\right]{\cal P}_{tc}\,,
\end{equation}
where
\begin{equation}\label{Ptc}
{\cal P}_{tc}\equiv\left|{\cal P}_{tc}\right|e^{i\delta{tc}}=
\left(P_t-P_c\right)+(P_{\rm ew}^t-P_{\rm ew}^c)
\end{equation}
and
\begin{equation}\label{rho-def}
\rho\,e^{i\theta}=\frac{\lambda^2R_b}{1-\lambda^2/2}
\left[1-\left(\frac{{\cal P}_{uc}+{\cal A}}{{\cal P}_{tc}}\right)\right].
\end{equation}
In these expressions, $\delta_{tc}$ and $\theta$ denote CP-conserving strong
phases, ${\cal P}_{uc}$ is defined in analogy to Eq.\ (\ref{Ptc}), 
$\lambda\equiv|V_{us}|=0.22$, $A\equiv\left|V_{cb}\right|/\lambda^2=
0.81\pm0.06$, and $R_b\equiv\left|V_{ub}/(\lambda V_{cb})\right|=0.36\pm0.08$.
The quantity $\rho\,e^{i\theta}$ is a measure of the strength of certain
rescattering effects, as will be discussed in more detail in 
Section~\ref{Sec:resc}.

If we apply the $SU(2)$ isospin symmetry of strong interactions, implying 
\begin{equation}
\tilde P_c=P_c\quad\mbox{and}\quad\tilde P_t=P_t\,, 
\end{equation}
the QCD penguin topologies with internal top and charm quarks contributing 
to $B^+\to\pi^+K^0$ and $B_d^0\to\pi^-K^+$ can be related to each other, 
yielding the following amplitude relations (for a detailed discussion, 
see Ref.\ 10):
\begin{eqnarray}
A(B^+\to\pi^+K^0)&\equiv&P\label{ampl-p}\\
A(B_d^0\to\pi^-K^+)&=&-\,\left[P+T+P_{\rm ew}\right],\label{ampl-n}
\end{eqnarray}
which play a central role to probe the CKM angle $\gamma$. Here the ``penguin''
amplitude $P$ is {\it defined} by the $B^+\to\pi^+K^0$ decay amplitude, the
quantity
\begin{eqnarray}
\lefteqn{P_{\rm ew}\equiv-\,|P_{\rm ew}|e^{i\delta_{\rm ew}}=-\left(1-
\frac{\lambda^2}{2}\right)\lambda^2A}\nonumber\\
&&\qquad\times\left[\left(\tilde P_{\rm ew}^t-\tilde
P_{\rm ew}^c\right)-\left(P_{\rm ew}^t-P_{\rm ew}^c\right)\right]
\end{eqnarray}
is essentially due to electroweak penguins, and 
\begin{eqnarray}
\lefteqn{T\equiv|T|e^{i\delta_T}e^{i\gamma}=\lambda^4A\,R_b\left[
\tilde{\cal T}-{\cal A}+\left(\tilde P_u-P_u\right)
\right.}\qquad\mbox{}\nonumber\\
&&\left.+\left(\tilde
P_{\rm ew}^u-\tilde P_{\rm ew}^t\right)-\left(P_{\rm ew}^u-
P_{\rm ew}^t\right)\right]e^{i\gamma}\qquad\mbox{}\label{T-def}
\end{eqnarray}
is usually referred to as a ``tree'' amplitude. However, owing to a subtlety 
in the implementation of the isospin symmetry, the amplitude $T$ does not only 
receive contributions from colour-allowed tree-diagram-like topologies, but 
also from penguin and annihilation topologies.\,\cite{bfm,defan}\, It is
an easy exercise to convince oneself that the amplitudes $P$, $T$ and 
$P_{\rm ew}$ are well-defined physical quantities.\,\cite{defan}

\begin{figure}
\centerline{
\rotate[r]{
\epsfxsize=6.5truecm
\epsffile{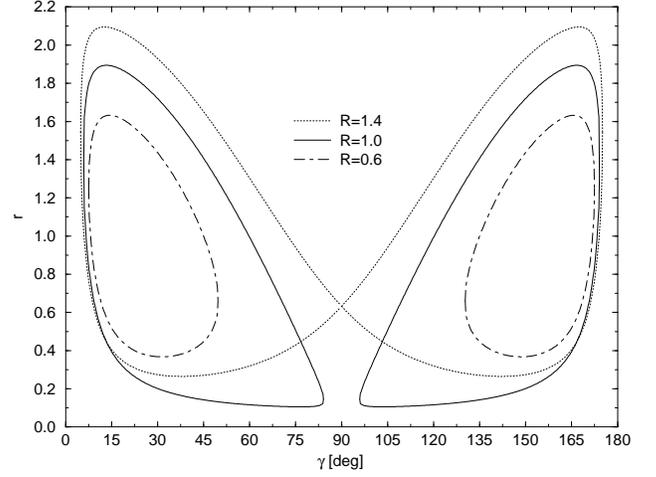}}}
\caption{The contours in the $\gamma\,$--$\,r$ plane for $|A_0|=0.2$ and for
various values of $R$ in the case of $\rho=\epsilon=0$.}\label{fig:contours}
\end{figure}

In the parametrization of the $B^\pm\to \pi^\pm K$ and $B_d\to\pi^\mp K^\pm$
observables, it turns out to be very useful to introduce the quantities
\begin{equation}
r\equiv\frac{|T|}{\sqrt{\langle|P|^2\rangle}}\,,\quad\epsilon\equiv
\frac{|P_{\rm ew}|}{\sqrt{\langle|P|^2\rangle}}\,,
\end{equation}
with $\langle|P|^2\rangle\equiv(|P|^2+|\overline{P}|^2)/2$, as well
as the CP-conserving strong phase differences
\begin{equation}
\delta\equiv\delta_T-\delta_{tc}\,,\quad\Delta\equiv\delta_{\rm ew}-
\delta_{tc}\,.
\end{equation}
In addition to the ratio $R$ of combined $B\to\pi K$ branching ratios defined 
by Eq.\ (\ref{Def-R}), also the ``pseudo-asymmetry'' 
\begin{equation}
A_0\equiv\frac{\mbox{BR}(B^0_d\to\pi^-K^+)-\mbox{BR}(\overline{B^0_d}\to
\pi^+K^-)}{\mbox{BR}(B^+\to\pi^+K^0)+\mbox{BR}(B^-\to\pi^-\overline{K^0})}
\end{equation}
plays an important role to probe the CKM angle $\gamma$. Explicit expressions
for $R$ and $A_0$ in terms of the parameters specified above are given
in Ref.\ 16.

\begin{figure}
\centerline{
\rotate[r]{
\epsfxsize=3truecm
\epsffile{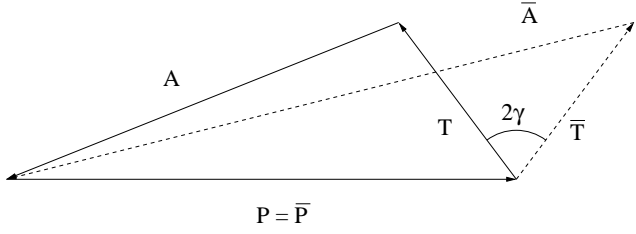}}}
\caption{Triangle construction with $A\equiv A(B^0_d\to\pi^-K^+)$ and 
$\overline{A}\equiv A(\overline{B^0_d}\to\pi^+K^-)$ to determine the CKM 
angle $\gamma$ in the case of $\rho=\epsilon=0$, where 
$P\equiv A(B^+\to\pi^+K^0)=A(B^-\to\pi^-\overline{K^0})\equiv
\overline{P}$.}\label{fig:triangles}
\end{figure}

\boldmath
\section{Strategies to Constrain and Determine the CKM Angle $\gamma$ with
the Help of $B^\pm\to\pi^\pm K$ and $B_d\to\pi^\mp K^\pm$ Decays}\label{strat}
\unboldmath
The observables $R$ and $A_0$ provide valuable information about the CKM
angle $\gamma$. If in addition to $R$ also the asymmetry $A_0$ can be 
measured, it is possible to eliminate the strong phase $\delta$ in the 
expression for $R$, and contours in the $\gamma\,$--$\,r$ plane can be 
fixed;\,\cite{defan} these are shown in Fig.\ \ref{fig:contours} for 
$|A_0|=0.2$ and for various values of $R$. These contours correspond to a 
mathematical implementation of a simple triangle construction,\,\cite{PAPIII} 
which is illustrated in Fig.\ \ref{fig:triangles}. In both  
Figs.\ \ref{fig:contours} and \ref{fig:triangles}, rescattering and EW 
penguin effects have been neglected for simplicity. A detailed study of 
their impact can be found in Refs.\ 16 and 17.

In order to determine the CKM angle $\gamma$, the quantity $r$, i.e.\ the 
magnitude of the ``tree'' amplitude $T$, has to be fixed. At
this step, a certain model dependence enters. In recent studies based on 
``factorization'', the authors of Refs.\ 3 and 4 came to the 
conclusion that a future theoretical uncertainty of $r$ as small as 
${\cal O}(10\%)$ may be achievable. In this case, the determination of 
$\gamma$ at future $B$-factories would be limited by statistics rather than 
by the uncertainty introduced through $r$, and $\Delta\gamma$ at the level 
of $10^\circ$ could in principle be achieved. However, since the properly 
defined amplitude $T$ (see Eq.\ (\ref{T-def})) does not only receive 
contributions from colour-allowed ``tree'' topologies, but also from 
penguin and annihilation processes,\,\cite{bfm,defan} it may be shifted 
sizeably from its ``factorized'' value so that $\Delta r={\cal O}(10\%)$ 
may be too optimistic. 

Interestingly, it is possible to derive bounds on $\gamma$ that do {\it not}
depend on $r$ at all.\,\cite{fm2}\, To this end, we eliminate again the
strong phase $\delta$ in the ratio $R$ of combined $B\to\pi K$ branching
ratios. If we now treat $r$ as a ``free'' variable, while keeping 
($\rho$, $\theta$) and ($\epsilon$, $\Delta$) fixed, we find that $R$ takes 
the following minimal value:\,\cite{defan} 
\begin{equation}\label{Rmin}
R_{\rm min}=\kappa\,\sin^2\gamma\,+\,
\frac{1}{\kappa}\left(\frac{A_0}{2\,\sin\gamma}\right)^2.
\end{equation}
In this expression, which is valid {\it exactly}, rescattering and EW 
penguin effects are described by
\begin{equation}\label{kappa-def}
\kappa=\frac{1}{w^2}\left[\,1+2\,(\epsilon\,w)\cos\Delta+
(\epsilon\,w)^2\,\right],
\end{equation}
with
\begin{equation}
w=\sqrt{1+2\,\rho\,\cos\theta\cos\gamma+\rho^2}.
\end{equation}
An allowed range for $\gamma$ is related to $R_{\rm min}$, since values of
$\gamma$ implying $R_{\rm exp}<R_{\rm min}$ are excluded ($R_{\rm exp}$ 
denotes the experimentally determined value of $R$). This range can
also be read off from the contour in the $\gamma\,$--$\,r$ plane 
corresponding to the measured values of $R$ and $A_0$, as can be seen in 
Fig.\ \ref{fig:contours}.   

The theoretical accuracy of these contours and of the associated bounds on 
$\gamma$ is limited by rescattering and EW penguin effects, which will be 
discussed in the following two sections. In the ``original'' bounds on 
$\gamma$ derived in Ref.\ 6, no information provided by $A_0$ has 
been used, i.e.\ both $r$ and $\delta$ were kept as ``free'' variables, 
and the special case $\rho=\epsilon=0$ has been assumed, implying 
$\sin^2\gamma<R_{\rm exp}$. Note that a measurement of $A_0\not=0$ allows 
us to exclude a certain range of $\gamma$ around $0^\circ$ and $180^\circ$.

\begin{figure}
\centerline{
\epsfxsize=6.85truecm
\epsffile{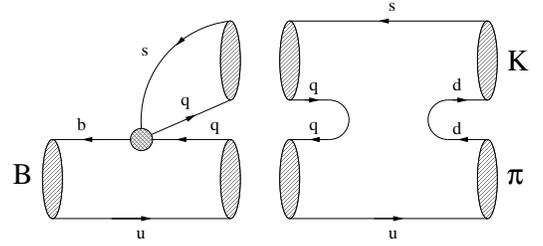}}
\caption{Illustration of rescattering processes of the kind $B^+\to
\bigl\{F_q^{(s)}\bigr\}\to\pi^+K^0$. The shaded circle represents insertions 
of the usual current--current operators $Q_{1,2}^q$ 
($q\in\{c,u\}$).}\label{fig:rescatter}
\end{figure}

\section{The Role of Rescattering Processes}\label{Sec:resc}
In the formalism discussed above, rescattering processes are closely related
to the quantity $\rho$ (see Eq.\ (\ref{rho-def})), which is highly 
CKM-suppressed by $\lambda^2R_b\approx0.02$ and receives contributions from 
penguin topologies with internal top, charm and up quarks, as well as from 
annihilation topologies. Na\"\i vely, one would expect that annihilation 
processes play a very minor role, and that penguins with internal top quarks 
are the most important ones. However, also penguins with internal charm and
up quarks lead, in general, to important contributions.\,\cite{LD-pens} Simple
model calculations, performed at the perturbative quark level, do not indicate
a significant compensation of the large CKM suppression of $\rho$ through
these topologies. However, these crude estimates do not take into account 
certain rescattering processes,\,\cite{FSI}$^{\mbox{-}}$\cite{atso} which 
may play an important role and can be divided into two classes:\,\cite{bfm}
\begin{itemize}
\item[i)] $B^+\to\{\overline{D^0}D_s^+,\,\overline{D^0}D_s^{\ast+},
\,\ldots\}\to\pi^+K^0$
\item[ii)]$B^+\to\{\pi^0K^+,\,\pi^0K^{\ast +},\,\ldots\}\to\pi^+K^0$,
\end{itemize}
where the dots include also intermediate multibody states. These processes 
are illustrated in Fig.\ \ref{fig:rescatter}. Here the shaded circle 
represents insertions of the usual current--current operators 
\begin{equation}\label{CC-def}
\begin{array}{rcl}
Q_1^q&=&(\bar q_{\alpha} s_{\beta})_{{\rm V-A}}
\;(\bar b_{\beta} q_{\alpha})_{{\rm V-A}}\\
Q_2^q&=&(\bar q_{\alpha} s_{\alpha})_{{\rm V-A}}\;
(\bar b_{\beta} q_{\beta})_{{\rm V-A}}\,, 
\end{array}
\end{equation}
where $\alpha$ and $\beta$ are colour indices, and $q\in\{c,u\}$. The
rescattering processes (i) and (ii) correspond to $q=c$ and $u$, respectively.

\begin{figure}
\centerline{
\rotate[r]{
\epsfxsize=6.5truecm
\epsffile{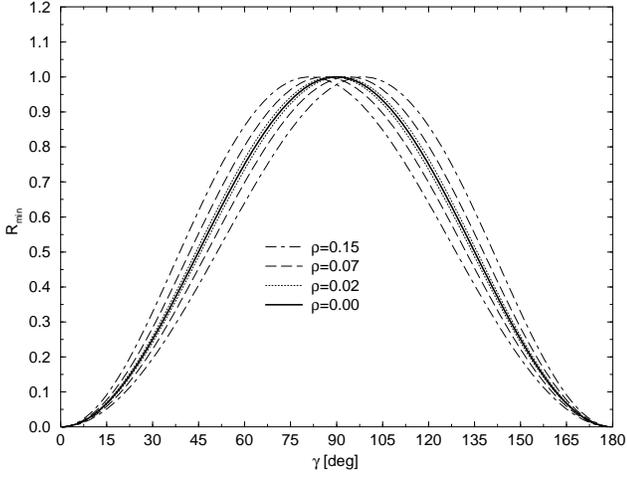}}}
\caption{The effect of final-state interactions on $R_{\rm min}$ for $A_0=0$.
The curves for a given value of $\rho$ correspond to 
$\theta\in\{0^\circ,180^\circ\}$ and represent the maximal shift from 
$\rho=0$.}\label{fig:RminFSI}
\end{figure}

If we look at Fig.\ \ref{fig:rescatter}, we observe that the 
final-state-interaction (FSI) effects of type (i) can be 
considered as long-distance 
contributions to penguin topologies with internal charm quarks, i.e.\ to the 
$P_c$ amplitude. They may affect BR$(B^\pm\to\pi^\pm K)$ significantly. 
On the other hand, the rescattering processes characterized by (ii) result in 
long-distance contributions to penguin topologies with internal up quarks 
and to annihilation topologies, i.e.\ to the amplitudes $P_u$ and
${\cal A}$. They play a minor role for BR$(B^\pm\to\pi^\pm K)$, but may 
affect assumption (ii) listed in Section~\ref{intro}, thereby leading to a 
sizeable CP asymmetry, $A_+$, as large as ${\cal O}(10\%)$ in this 
mode.\,\cite{gewe}$^{\mbox{-}}$\cite{atso}\, The point is as follows: while 
we would have $\rho\approx0$ if rescattering processes of type (i) played the 
dominant role in $B^+\to\pi^+ K^0$, or $\rho={\cal O}(\lambda^2R_b)$ if both 
processes had similar importance, $\rho$ would be as large as 
${\cal O}(10\%)$ if the FSI effects characterized by (ii) would dominate 
$B^+\to\pi^+ K^0$ so that $|{\cal P}_{uc}|/|{\cal P}_{tc}|={\cal O}(5)$. 
This order of magnitude is found in a recent attempt to evaluate rescattering
processes of the kind $B^+\to\{\pi^0K^+\}\to\pi^+K^0$ with the help of 
Regge phenomenology.\,\cite{fknp}\, A similar feature is also present in 
other approaches to deal with these FSI effects.\,\cite{gewe,neubert}\,  
Therefore, we have arguments that rescattering processes may play an 
important role. 

A detailed study of their impact on the constraints on $\gamma$ arising from 
the $B^\pm\to \pi^\pm K$ and $B_d\to\pi^\mp K^\pm$ observables was performed 
in Ref.\ 16. While these effects, which are included in the 
formalism discussed above through the parameter $\kappa$ (see Eq.\ 
(\ref{kappa-def})), are minimal for $\theta\in\{90^\circ,270^\circ\}$ and 
only of second order, they are maximal for $\theta\in\{0^\circ,180^\circ\}$. 
In Fig.\ \ref{fig:RminFSI}, these maximal effects are shown for various values 
of $\rho$ in the case of $A_0=0$. Looking at this figure, we observe that we 
have negligibly small effects for $\rho=0.02$, which was assumed in 
Ref.\ 6 in the form of point~(ii) listed in Section~\ref{intro}. 
For values of $\rho$ as large as $0.15$, we have an uncertainty for 
$\gamma_0$ (see Eqs.\ (\ref{gamma-bound1}) and (\ref{gam0})) of at most 
$\pm10^\circ$. 

\begin{figure}
\centerline{
\rotate[r]{
\epsfxsize=6.5truecm
\epsffile{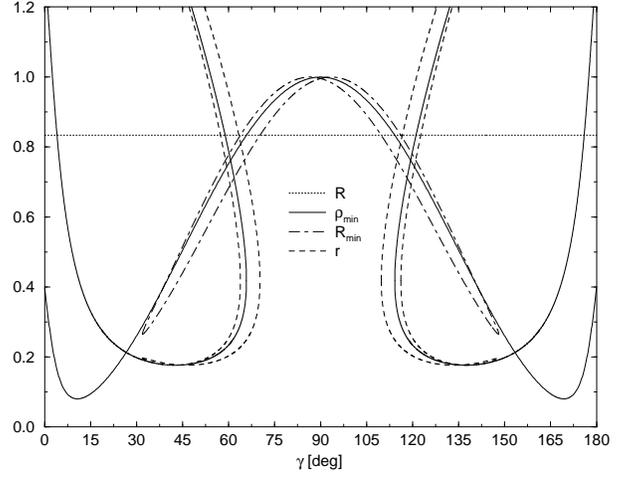}}}
\caption{Illustration of the strategy to control rescattering effects in 
$R_{\rm min}$ and the contours in the $\gamma\,$--$\,r$ plane through
the decay $B^\pm\to K^\pm K$ within a simple model (for details, see
Ref.\ 17).}\label{fig:FSIcontrol}
\end{figure}

The FSI effects can be controlled through experimental data. A first step 
towards this goal is provided by the CP asymmetry $A_+$. It implies 
an allowed range for $\rho$, which is given by $\rho_{\rm min}\,\leq\,\rho\,
\leq\,\rho^{\rm max}$, with 
\begin{equation}\label{rho-min-max}
\rho_{\rm min}^{\rm max}=\frac{\sqrt{A_+^2+
\left(1-A_+^2\right)\sin^2\gamma}\pm\sqrt{\left(1-A_+^2\right)
\sin^2\gamma}}{|A_+|}.
\end{equation}
In order to go beyond these constraints, $B^\pm\to K^\pm K$ decays -- the 
$SU(3)$ counterparts of $B^\pm\to \pi^\pm K$ -- play a key role, allowing us 
to include the rescattering processes in the contours in the 
$\gamma\,$--$\,r$ plane and the associated constraints on $\gamma$ completely,
as was pointed out in Refs.\ 16 and 17 (for alternative strategies, see 
Refs.\ 10 and 14). As a by-product, this strategy moreover gives  
an allowed region for $\rho$, and excludes values of $\gamma$ within ranges 
around $0^\circ$ and $180^\circ$. It is interesting to note that $SU(3)$ 
breaking enters in this approach only at the ``next-to-leading order'' 
level, as it represents a correction to the correction to the bounds on 
$\gamma$ arising from rescattering processes. Moreover, this strategy also 
works if the CP asymmetry $A_+$ arising in $B^+\to\pi^+ K^0$ should turn out 
to be very small. In this case, there may also be large rescattering effects, 
which would then not be signalled by sizeable CP violation in this channel.

Following Ref.\ 17, this approach to control the FSI effects is 
illustrated in Fig.\ \ref{fig:FSIcontrol} by showing the contours in the 
$\gamma\,$--$\,r$ plane and the dependence of $R_{\rm min}$ on the CKM 
angle $\gamma$. Here the simple model advocated by the authors of 
Refs.\ 12 and 13 was used to obtain values for the $B\to\pi K$, $KK$ 
observables by choosing a specific set of input parameters (for details, 
see Ref.\ 17). The value of $R=0.83$ arising in this case is 
represented in Fig.\ \ref{fig:FSIcontrol} by the dotted line. It is an 
easy exercise to read off the corresponding allowed range for $\gamma$ from
this figure. 

Since the ``short-distance'' expectation for the combined branching ratio
BR$(B^\pm\to K^\pm K)$ is ${\cal O}(10^{-6})$,\,\cite{AKL} experimental 
studies of $B^\pm\to K^\pm K$ appear to be difficult. These modes have not 
yet been observed, and only upper limits for BR$(B^\pm\to K^\pm K)$ are 
available.\,\cite{cleo,newCLEO}\, However, rescattering effects may enhance 
this quantity significantly, and could thereby make $B^\pm\to K^\pm K$ 
measurable at future $B$-factories.\,\cite{defan,rf-FSI}\, Another important 
indicator of large FSI effects is provided by $B_d\to K^+K^-$ 
decays,\,\cite{groro-FSI} for which stronger experimental bounds already 
exist.\,\cite{cleo,newCLEO}\,

Although $B^\pm\to K^\pm K$ decays allow us to determine the shift of the 
contours in the $\gamma\,$--$\,r$ plane arising from rescattering processes, 
they do not allow us to take into account these effects also in the 
determination of $\gamma$, requiring some knowledge on $r$, in contrast 
to the bounds on $\gamma$. As we have already noted, this quantity is not 
just the ratio of a ``tree'' to a ``penguin'' amplitude, which is the usual 
terminology, but has a rather complex structure and may 
in principle be considerably affected by FSI effects. However, if future
measurements of BR$(B^\pm\to K^\pm K)$ and BR$(B_d\to K^+K^-)$ should not 
show a significant enhancement with respect to the ``short-distance'' 
expectations of ${\cal O}(10^{-6})$ and ${\cal O}(10^{-8})$, respectively,
and if $A_+$ should not be in excess of ${\cal O}(1\%)$, a future theoretical
accuracy of $r$ as small as ${\cal O}(10\%)$ may be achievable.

\begin{figure}
\centerline{
\rotate[r]{
\epsfxsize=6.5truecm
\epsffile{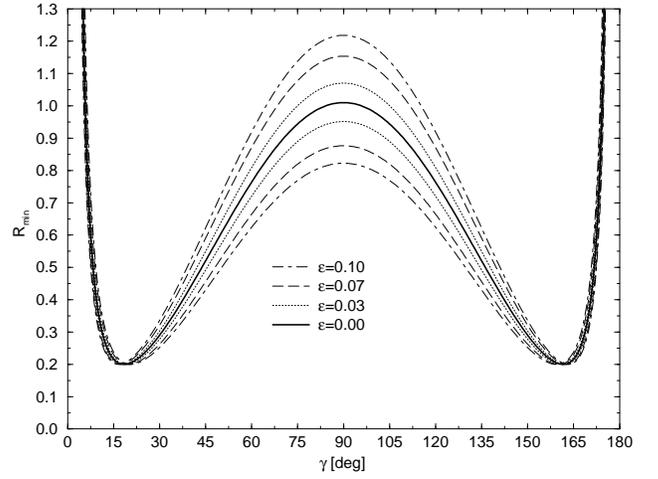}}}
\caption{The effect of electroweak penguins on $R_{\rm min}$ for $|A_0|=0.2$.
The curves for a given value of $\epsilon$ correspond to 
$\Delta\in\{0^\circ,180^\circ\}$ and represent the maximal shift from 
$\epsilon=0$.}\label{fig:RminEW}
\end{figure}

\section{The Role of Electroweak Penguins}

The modification of $R_{\rm min}$ through EW penguin topologies is described 
by $\kappa=1+2\,\epsilon\,\cos\Delta+\epsilon^2$. These effects are minimal
and only of second order in $\epsilon$ for $\Delta\in\{90^\circ,270^\circ\}$, 
and maximal for $\Delta\in\{0^\circ,180^\circ\}$. In the case of 
$\Delta=0^\circ$, which is favoured by ``factorization'', the bounds on 
$\gamma$ get stronger, excluding a larger region around $\gamma=90^\circ$, 
while they are weakened for $\Delta=180^\circ$. In Fig.~\ref{fig:RminEW}, 
the maximal EW penguin effects are shown for $|A_0|=0.2$ and for various 
values of $\epsilon$. The EW penguins are ``colour-suppressed'' in the case 
of $B^+\to\pi^+K^0$ and $B^0_d\to\pi^-K^+$; estimates based on simple 
calculations performed at the perturbative quark level, where
the relevant hadronic matrix elements are treated within the ``factorization''
approach, typically give $\epsilon={\cal O}(1\%)$.\,\cite{AKL}\, These 
crude estimates may, however, underestimate the role of these
topologies.\,\cite{groro,neubert}\, 

An improved theoretical description of the EW penguins is possible, using 
the general expressions for the corresponding four-quark operators and 
performing appropriate Fierz transformations. Following these 
lines,\,\cite{defan} we arrive at the expression
\begin{equation}\label{eps-r-final}
\frac{\epsilon}{r}\,e^{i(\Delta-\delta)}\approx\frac{3}{2\lambda^2R_b}\left[
\frac{C_1'(\mu)C_{10}(\mu)-C_2'(\mu)C_9(\mu)}{C_2'^2(\mu)-
C_1'^2(\mu)}\right]a\,e^{i\omega}
\end{equation}
with $C_1'(\mu)\equiv C_1(\mu)+3\,C_9(\mu)/2$ and $C_2'(\mu)\equiv C_2(\mu)+
3\,C_{10}(\mu)/2$, where $C_{1,2}(\mu)$ are the Wilson coefficients of the 
current--current operators specified in Eq.\ (\ref{CC-def}), and 
$C_{9,10}(\mu)$ those of the EW penguin operators 
\begin{equation}
\begin{array}{rcl}
Q_9&=&\frac{3}{2}(\bar b_{\alpha}s_{\alpha})_{{\rm V-A}}
\sum\limits_{q=u,d,c,s,b}c_q\,(\bar q_{\beta}q_{\beta})_{{\rm V-A}}\\
Q_{10}&=&\frac{3}{2}(\bar b_{\alpha}s_{\beta})_{{\rm V-A}}
\sum\limits_{q=u,d,c,s,b}c_q\,(\bar q_{\beta} q_{\alpha})_{{\rm V-A}}\,.
\end{array}
\end{equation}
The combination of Wilson coefficients in Eq.\ (\ref{eps-r-final}) is 
essentially renormalization-scale-independent and changes only by 
${\cal O}(1\%)$ when evolving from $\mu=M_W$ down to $\mu=m_b$. Employing 
$R_b=0.36$ and typical values for the Wilson coefficients yields\,\cite{defan}
\begin{equation}
\frac{\epsilon}{r}\,e^{i(\Delta-\delta)}\approx 0.75\times a\,e^{i\omega}.
\end{equation}
The quantity $a\,e^{i\omega}$ is given by 
\begin{equation}
a\,e^{i\omega}\equiv\frac{a_2^{\rm eff}}{a_1^{\rm eff}}\,,
\end{equation}
where $a_1^{\rm eff}$ and $a_2^{\rm eff}$ correspond to a generalization of
the usual phenomenological colour factors $a_1$ and $a_2$ describing the 
``strength'' of colour-suppressed and colour-allowed decay processes, 
respectively.\,\cite{defan}\, Comparing experimental data 
on $B^-\to D^{(\ast)0}\pi^-$ and $\overline{B^0_d}\to D^{(\ast)+}\pi^-$, 
as well as on $B^-\to D^{(\ast)0}\rho^-$ and $\overline{B^0_d}\to 
D^{(\ast)+}\rho^-$ decays gives $a_2/a_1={\cal O}(0.25)$, 
where $a_1$ and $a_2$ are -- in contrast to $a_1^{\rm eff}$ and 
$a_2^{\rm eff}$ -- real quantities, and their relative sign is found to be 
positive. For $a=0.25$, we obtain a value of $\epsilon/r$ that is larger
than the ``factorized'' result
\begin{equation}
\left.\frac{\epsilon}{r}\,e^{i(\Delta-\delta)}\right|_{\rm fact}=\,0.06
\end{equation}
by a factor of 3. A detailed study of the effects of the EW penguins described
by Eq.\ (\ref{eps-r-final}) on the strategies to probe the CKM angle $\gamma$ 
discussed in Section \ref{strat} was performed in Ref.\ 16. There it was also 
pointed out that a first step towards the experimental control of the 
``colour-suppressed'' EW penguin contributions to the $B\to\pi K$ amplitude 
relations (\ref{ampl-p}) and (\ref{ampl-n}) is provided by the decay 
$B^+\to\pi^+\pi^0$. More refined strategies will certainly be developed in 
the future, when better experimental data become available.

\boldmath
\section{Probing $\gamma$ with $B_s\to K\overline{K}$ Decays}
\unboldmath

In this section, we focus on the modes $B_s\to K^0\overline{K^0}$ and
$B_s\to K^+K^-$, which are the $B_s$ counterparts of the 
$B_{u,d}\to\pi K$ decays discussed above, where the up and down ``spectator'' 
quarks are replaced by a strange quark.  Because of the expected large 
$B^0_s$--$\overline{B^0_s}$ mixing parameter 
$x_s\equiv\Delta M_s/\Gamma_s={\cal O}(20)$, experimental studies of CP 
violation in $B_s$ decays are regarded as being very difficult. In 
particular, an excellent vertex resolution system is required to keep track 
of the rapid oscillatory $\Delta M_st$ terms arising in tagged $B_s$ decays. 
These terms cancel, however, in the untagged $B_s$ decay rates defined by
\begin{equation}\label{untagged}
\Gamma[f(t)]\equiv\Gamma(B_s^0(t)\to f)\,+\,\Gamma(\overline{B^0_s}(t)\to f)\,,
\end{equation}
where one does not distinguish between initially, i.e.\ at time $t=0$, 
present $B^0_s$ and $\overline{B^0_s}$ mesons. In this case, the expected 
sizeable width difference\,\cite{DG-calc} 
$\Delta\Gamma_s\equiv\Gamma_H^{(s)}-\Gamma_L^{(s)}$ 
between the mass eigenstates $B_s^H$ (``heavy'') and $B_s^L$ (``light'') of 
the $B_s$ system may provide an alternative route to explore CP 
violation.\,\cite{dunietz}\, Several strategies were proposed to extract 
CKM phases from experimental studies of such untagged $B_s$ 
decays.\,\cite{dunietz}$^{\mbox{-}}$\cite{fd2}\,

In Ref.\ 24, it was pointed out that the modes $B_s\to K^0\overline{K^0}$ 
and $B_s\to K^+K^-$ probe the CKM angle $\gamma$. Their decay 
amplitudes take a form completely analogous to Eqs.\ (\ref{ampl-p}) and 
(\ref{ampl-n}), and the corresponding untagged decay rates can be
expressed as follows:\,\cite{bsgam}
\begin{eqnarray}
\Gamma[K^0\overline{K^0}(t)]&=&R_L\,e^{-\Gamma_L^{(s)} t}
+R_H\,e^{-\Gamma_H^{(s)} t}\label{Bk0k0bar}\\
\Gamma[K^+K^-(t)]&=&\Gamma[K^0\overline{K^0}(0)]
\left[a\,e^{-\Gamma_L^{(s)} t}
+b\,e^{-\Gamma_H^{(s)} t}\right].\qquad\mbox{}\label{Bkpkm}
\end{eqnarray}
Since we have $a+b=R_s$, where $R_s$ corresponds to the ratio $R$ of the 
combined $B\to\pi K$ branching ratios (see Eq.~(\ref{Def-R})), 
bounds on $\gamma$ similar to those discussed in Sections~\ref{intro} and 
\ref{strat} can also be obtained from the untagged $B_s\to K\overline{K}$ 
observables. Moreover, a comparison of $R$ and $R_s$ provides valuable
insights into $SU(3)$ breaking. 

A closer look shows, however, that it is possible to derive 
more elaborate bounds from the untagged $B_s\to K\overline{K}$ 
rates:\,\cite{bsgam}\, 
\begin{equation}\label{cot-range}
\frac{\left|\,1-\sqrt{a}\,\right|}{\sqrt{b}}\leq|\cot\gamma\,|\leq\frac{1+
\sqrt{a}}{\sqrt{b}}\,,
\end{equation}
corresponding to the allowed range 
\begin{equation}\label{Bs-bounds}
\gamma_1\leq\gamma\leq\gamma_2\quad\lor\quad180^\circ-\gamma_2\leq\gamma
\leq180^\circ-\gamma_1
\end{equation}
with
\begin{equation}
\gamma_1\equiv\mbox{arccot}\left(\frac{1+\sqrt{a}}{\sqrt{b}}\right),\quad
\gamma_2\equiv\mbox{arccot}\left(\frac{\left|\,1-\sqrt{a}\,\right|}{\sqrt{b}}
\right).
\end{equation}
Besides a sizeable value of $\Delta\Gamma_s$ and non-vanishing observables 
$a$ and $b$, the bound (\ref{cot-range}) does not require any constraint on 
these observables such as $R_s=a+b<1$, which is needed for Eqs.\ 
(\ref{gamma-bound1}) and (\ref{gam0}) to become effective.

As in the $B\to\pi K$ case, the theoretical accuracy of these constraints, 
which make use only of the general amplitude structure arising within the 
Standard Model and of the $SU(2)$ isospin symmetry of strong interactions, 
is also limited by certain rescattering processes and contributions arising 
from EW penguins. In Eq.\ (\ref{cot-range}), these effects are neglected
for simplicity. The completely general formalism, taking also into account 
these effects, is derived in Ref.\ 26, where also strategies to
control them through experimental data are discussed. 

\begin{figure}
\centerline{
\rotate[r]{
\epsfxsize=6.5truecm
\epsffile{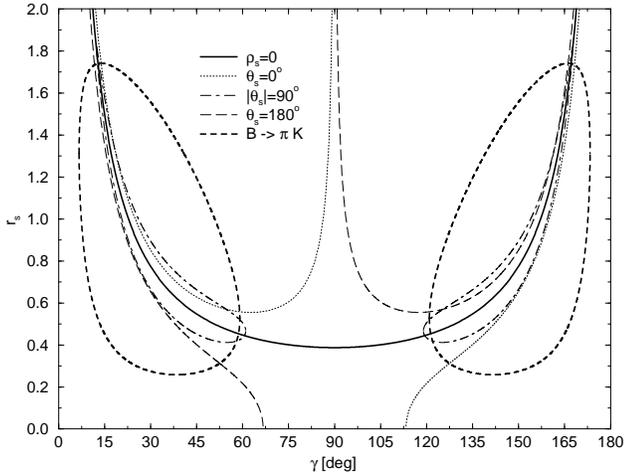}}}
\caption{The contours in the $\gamma\,$--$\,r_{(s)}$ plane for 
$a=0.60$, $b=0.15$ and $\rho_s=0.15$, $\epsilon_s=0$. The 
$B\to\pi K$ contours correspond to $R=0.75$, $A_0=0.2$ and 
$\rho=\epsilon=0$.}\label{fig:r-cont}
\end{figure}

In order to go beyond these constraints and to {\it determine} $\gamma$ from 
the untagged $B_s\to K\overline{K}$ observables, the magnitude of an amplitude
$T_s$, which corresponds to $T$ (see Eq.\ (\ref{T-def})), has to be fixed,
leading to hadronic uncertainties similar to those in the $B\to\pi K$ case.
Such an input can be avoided by considering the contours in the
$\gamma\,$--$\,r_{(s)}$ and $\gamma\,$--$\,\cos\delta_{(s)}$ planes, and 
applying the $SU(3)$ flavour symmetry to relate $r_s$ to $r$ and 
$\cos\delta_s$ to $\cos\delta$, respectively.\,\cite{bsgam} The contours in 
the $\gamma\,$--$\,r_{(s)}$ plane are illustrated in Fig.~\ref{fig:r-cont}. 
Using the formalism presented in Refs.\ 16, 17 and 26, rescattering and 
EW penguin effects can be included in these contours. As a ``by-product'', 
also values for the hadronic quantities $r_{(s)}$ and $\cos\delta_{(s)}$
are obtained, which are of special interest to test the factorization 
hypothesis.

Provided a tagged, time-dependent measurement of $B_s\to K^0\overline{K^0}$ 
and $B_s\to K^+K^-$ can be performed,\,\cite{kly} it would be 
possible to extract $\gamma$ in such a way that rescattering effects are
taken into account ``automatically''.\,\cite{bsgam}\, To this end, the 
$B_s\to K\overline{K}$ observables are sufficient, and the theoretical 
accuracy of $\gamma$ would only be limited by EW penguins. Let me finally 
note that the $B_s\to K\overline{K}$ decays represent also an interesting 
probe for certain scenarios of physics beyond the Standard Model.\,\cite{bsgam}
 
\section{Conclusions}

On the long and winding road towards the extraction of the CKM angle $\gamma$, 
the decays $B^\pm\to \pi^\pm K$ and $B_d\to\pi^\mp K^\pm$ are expected to
play an important role. An accurate measurement of these modes, as well 
as of $B\to KK$ and $B\to\pi\pi$ decays to control rescattering and EW 
penguin effects, is therefore an important goal of the future $B$-factories. 
At present, data for these decays are already starting to become available,
and the coming years will certainly be very exciting. The modes 
$B_s\to K^0\overline{K^0}$ and $B_s\to K^+K^-$ also offer interesting 
strategies to probe the CKM angle $\gamma$. Here the width difference 
$\Delta\Gamma_s$ may provide an interesting tool to accomplish this task. 
In order to investigate $B_s$ decays, experiments at hadron machines 
appear to be most promising.

\end{document}